\documentclass[]{spie}  

\usepackage{amsmath,amsfonts,amssymb}
\usepackage{graphicx}
\usepackage[colorlinks=true, allcolors=blue]{hyperref}

\title{Wavefront sensor for millimeter/submillimeter-wave adaptive optics based on aperture-plane interferometry}

\author[a]{Yoichi Tamura}
\author[b]{Ryohei Kawabe}
\author[c]{Yuhei Fukasaku}
\author[a]{Kimihiro Kimura}
\author[a]{Tetsutaro Ueda}
\author[a]{Akio Taniguchi}
\author[d]{Nozomi Okada}
\author[d]{Hideo Ogawa}
\author[e]{Ikumi Hashimoto}
\author[b,f]{Tetsuhiro Minamidani}
\author[b]{Noriyuki Kawaguchi}
\author[c]{Nario Kuno}
\author[a]{Yohei Togami}
\author[a]{Masato Hagimoto}
\author[a]{Satoya Nakano}
\author[a]{Keiichi Matsuda}
\author[g]{Sachiko Okumura}
\author[g]{Tomoko Nakamura}
\author[h]{Mikio Kurita}
\author[j,i]{Tatsuya Takekoshi}
\author[b]{Tai Oshima}
\author[d]{Toshikazu Onishi}
\author[i]{Kotaro Kohno}
\affil[a]{Nagoya University, Nagoya, 464-8602 Japan}
\affil[b]{National Astronomical Observatory of Japan, Mitaka, Tokyo, 181-8588 Japan}
\affil[c]{University of Tsukuba, Tsukuba, Ibaraki, 305-8573 Japan}
\affil[d]{Dept.\ Physical Science, Osaka Prefecture University, Sakai, Osaka, 599-8531 Japan}
\affil[e]{Dept.\ Aerospace Engineering, Osaka Prefecture University, Sakai, Osaka, 599-8531 Japan}
\affil[f]{Nobeyama Radio Observatory, Nagano, 384-1305 Japan}
\affil[g]{Japan Women's University, Bunkyo, Tokyo, 112-8681 Japan}
\affil[h]{Kyoto University, Kyoto, 606-8501 Japan}
\affil[i]{Institute of Astronomy, The University of Tokyo, Mitaka, Tokyo, 181-0015 Japan}
\affil[j]{Kitami Institute of Technology, Kitami, Hokkaido, 090-8507 Japan}

\authorinfo{Further author information: (Send correspondence to Y.T.)\\Y.T.: E-mail: ytamura@nagoya-u.jp, Telephone: +81 (0)52 789 2846}

\begin{document} 
\maketitle

\begin{abstract}
We present a concept of a millimeter wavefront sensor that allows real-time sensing of the surface of a ground-based millimeter/submillimeter telescope. It is becoming important for ground-based millimeter/submillimeter astronomy to make telescopes larger with keeping their surface accurate. To establish `millimetric adaptive optics (MAO)' that instantaneously corrects the wavefront degradation induced by deformation of telescope optics, our wavefront sensor based on radio interferometry measures changes in excess path lengths from characteristic positions on the primary mirror surface to the focal plane. This plays a fundamental role in planed 50-m class submillimeter telescopes such as LST and AtLAST.
\end{abstract}


\section{INTRODUCTION}
\label{sec:intro}

The increase in size of telescopes, i.e., the acquisition of larger collecting area and spatial resolution, is the very history of observational astronomy. For millimeter and submillimeter-wave telescopes (mm/submm telescopes, hereafter), increasing the antenna diameter with keeping the mirror surface accurate is important in developing new astronomical fields.
The advent of an aperture synthesis interferometer has resulted in significant improvement in collecting area and angular resolution. For example, the Atacama Large Millimeter/submillimeter Array (ALMA) realized the performance by keeping the size of the array element antennas moderate ($D = 7$ and 12~m) and increasing stiffness of them to keep their surface down to $20$~$\mu$m r.m.s or even better. 

However, aperture synthesis interferometry is not a panacea in realizing a larger-scale mm/submm telescope. The focal plane instruments of radio interferometers are limited to coherent receivers (e.g., heterodyne receivers) that can detect the phase of celestial signals. On the other hand, large-format arrays of direct photon detectors (e.g., cameras and integrated superconducting spectrometers\cite{Endo19}), which have prospered in recent years, cannot be accommodated as receivers for an aperture synthesis interferometer, because phase detection is not possible. Thus, it becomes increasingly important to have a large-aperture single-dish mm/submm telescope (e.g., Large Submillimeter Telescope\cite{Kawabe16} and Atacama Large Aperture Submillimeter Telescope\cite{Klaassen20}), especially in the field in which direct photon detector arrays play a crucial role. 
%

A limiting factor in size and operating frequency of mm/submm telescopes is the deterioration of optical performance due to changes in the environment surrounding the telescope, such as wind load and thermal deformation in addition to gravitational deformation.\footnote{In the millimeter and submillimeter regime, the inhomogeneities in column density of water vapor in the troposphere also cause wavefront degradation. However, the spatial scale at which the water vapor fluctuations dominate the wavefront degradation is typically much longer than the telescope's aperture, for example, $\gtrsim 10^2$~m at the ALMA site. For this reason, in most cases, the Earth's atmosphere is not a major source of wavefront degradation for single-dish mm/submm telescopes.}
This is unlike the visible/near-infrared (NIR) regime, where the major cause of wavefront degradation is due to temperature inhomogeneities of the tropospheric atmosphere on the order of tens of cm.  Mm/submm telescopes have been manufactured on the premise of large ($> 10$~m) structures in order to ensure the large collecting area and high spatial resolution. For this reason, building a huge dome to cover it would require enormous costs. Therefore, it was an issue to construct an `exposed' antenna outside and ensure its mirror surface with an accuracy of $\sim 20$~$\mu$m.

With the existing technology, it is possible to measure the mirror surface shape in advance by the photogrammetry or radio holography methods, and to correct the mirror surface down to the level of $\sim 10$~$\mu$m (r.m.s.) by adjusting the positions of the primary mirror panels. 
In recent years, the weight reduction and enhanced stiffness of the primary mirror structure and the active surface control of the primary mirror panels by the motorized actuators have been used to keep the primary mirror an ideal shape with respect to changes in telescope elevation angle.\cite{Hughes10}

The problem is, however, the deformation of the primary mirror surface and the secondary mirror support structure due to wind load and thermal deformation. This is because the time-scale of the deformation is fairly short compared with a typical time-scale of astronomical observations. It is difficult to measure the mirror surface shape in real time during the observations. It is likely that most of the deformation across the primary mirror are dominated by lower-order deformation modes. Their typical spatial scale is found to be a fraction of the aperture diameter (1--10~m)\cite{Levy96}. The time-scale of the deformation is determined by the natural frequency of the primary mirror structure to the wind load and is typically $\sim 10^{-1}$--1~s (e.g., $\simeq 1.1$~s for the Nobeyama 45~m mm-wave telescope\cite{Smith00}).
In this case, the control technology of the adaptive optics does not become a major obstacle. The spatial and time-scales of adaptive optics control, which is already established in NIR astronomy, are on the order of $\sim 10$~cm and $\sim 1$~kHz, respectively, which are several orders of magnitude higher in spatial and temporal frequencies than those of large mm/submm telescopes.
Therefore, in the radio telescope, the wavefront compensation can be performed at the primary or secondary mirror equipped with the active surface control, instead of a high-speed deformable mirror employed in NIR telescopes.

Then, the major obstacle in the mm/submm regime lies in \emph{how to measure the wavefront in real time}.
Wavefront sensing using a Shack-Hartmann sensor is the mainstream in the optical and NIR. A similar approach, however, is not applicable to MAO, since no large-format detector array that can be manufactured cheaply is available in the mm/submm.  Instead, radio astronomy can exploit radio interferometry as a native wavefront sensing technology, which measures the difference between the arrival times of the wavefronts coming through two independent optical paths.

In this paper, we present a concept of a mm wavefront sensor that allows real-time sensing of the primary mirror surface of a ground-based large-aperture mm/submm telescope.


\begin{figure}
    \centering
    \includegraphics[width=1.0\textwidth]{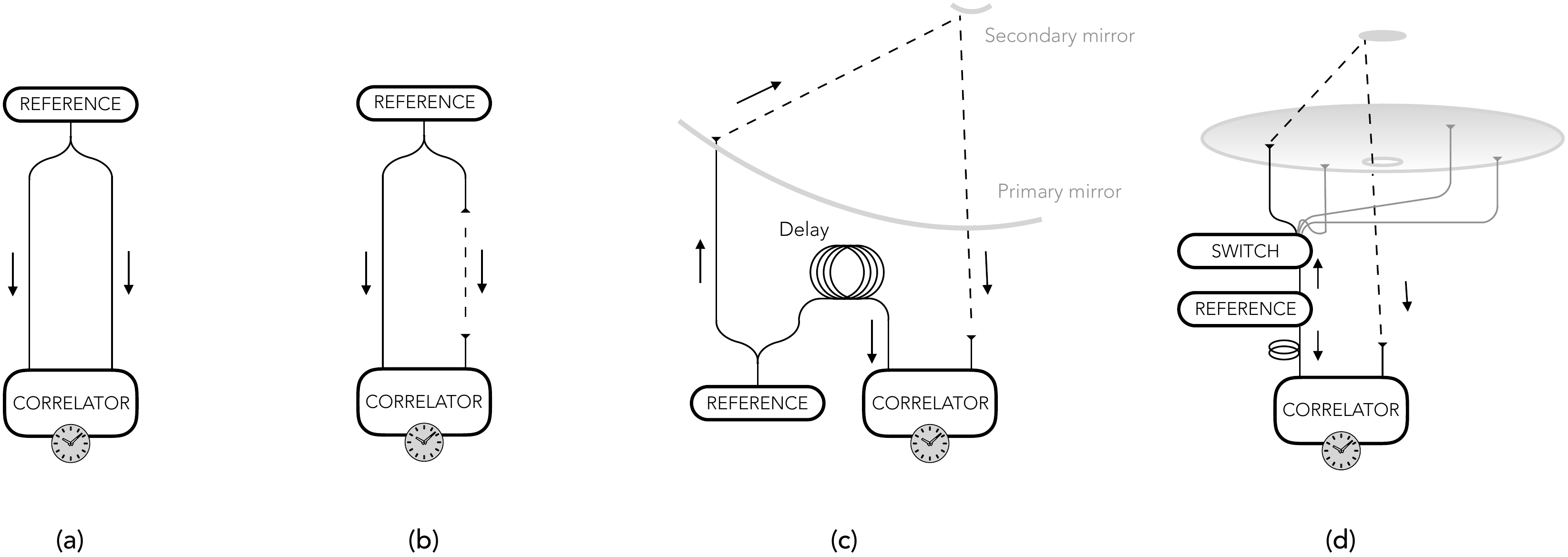}
    \caption{A schematic diagram of aperture-plane interferometry. (a) A simple interferometer with a correlator which measures a difference in arrival time between two signals coming from a common reference source. (b) The same as (a) but one of the optical paths coming through the free space as shown with the dashed line. (c) The same as (b) but the path through the free space goes through the telescope optics. The lengths of the other path is adjusted by a delay line shown as a loop. (d) The same as (c) with a switch followed by an array of radiators which enables to measure the distances from arbitrary positions to the focal point.}
    \label{fig:fig_api}
\end{figure}


\section{APERTURE-PLANE INTERFEROMETRY FOR WAVEFRONT SENSING}
\label{sec:principle}

As seen in the previous section, what needs to be measured first is the relative time variation of the deformation of the optical system, rather than wavefronts of celestial signals coming through the atmosphere. This is equivalent to the errors in excess path length (EPL) from the focus to arbitrary points on the primary surface.
Therefore, the wavefront sensor employs multiple reference microwave/mm sources placed on the aperture of the telescope as a phase standard. The goal of the surface accuracy we measure with a prototype sensor (see later sections) is 40~$\mu$m r.m.s. This is close to $\approx 100~\mu$m r.m.s., which is routinely achieved by holographic measurements of the Nobeyama 45~m mm-wave telescope.  This measurement accuracy is equivalent to the phase accuracy of 1$^\circ$ r.m.s.\ when operating the wavefront sensor at 20~GHz.

Figure~\ref{fig:fig_api} shows the schematic diagram representing the principle of aperture-plane interferometry. A simple interferometer comprising a correlator (figure~\ref{fig:fig_api}a, b) allows us to measure a difference in arrival time between two signals generated by a common reference source. The interferometer in which one of the paths goes through the telescope optics (figure~\ref{fig:fig_api}c) measures the EPL from a certain point on the primary surface to the focus of the telescope.  This is similar to what is used for phase calibration (the p-cal method, REF) of phase-referencing receivers used in VLBI Exploration of Radio Astronomy (VERA\cite{Kobayashi03}).  The method is a scalable technique and can be time-multiplexed by an intervening switch which is followed by a series of radiators placed across the surface of the primary mirror (figure~\ref{fig:fig_api}d).

The required number of reference points (i.e., radiators) depends on to what order of deformation modes needs to be characterized. As the wavefront sensor detects the relative deformation with respect to the ideal surface realized when the telescope structure is static, the surface needs to be adjusted by a conventional method such as the radio holography in advance. The relative deformation of the structure induced by wind and thermal loads emerges mostly on large spatial scales (e.g., a fraction of an aperture size).\cite{Levy96} Therefore, a couple of tens references across the aperture will be good enough.

We opt broadband noise as the common reference source, rather than a continuous wave (CW) or narrowband signals. This is because in the case of CW or narrowband signals, the stray lights through multiple paths can interfere with each other, resulting in considerable systematic errors in measurements of EPLs. On the other hand, broadband noises are more robust against the multi-path interference, and thus they are often used for delay calibration in aperture-synthesis interferometers such as ALMA.
The EPL is imprinted as a phase slope in a cross power spectrum (CPS) of the broadband noise. Let $C(\nu;\,\mathrm{EPL}) = A(\nu)\exp{[i\phi(\nu;\,\mathrm{EPL})]}$ be a measured CPS. Then, a calibrated CPS is obtained as $C(\nu;\,\mathrm{EPL})/C(\nu; 0) = \exp{[i\{\phi(\nu;\,\mathrm{EPL})-\phi(\nu;\,0)\}]}$. The phase of a calibrated CPS, $\Delta\phi(\nu;\,\mathrm{EPL}) \equiv \phi(\nu;\,\mathrm{EPL})-\phi(\nu;\,0)$, induced by a certain EPL is expressed as a function of frequency $\nu$ as
\begin{eqnarray}
\Delta\phi(\nu) = 2\pi\tau\nu = \frac{2\pi\,\mathrm{EPL}}{c}\nu ,
\end{eqnarray}\label{eq:epl}
where $\tau$ is the time delay induced by EPL. This means that the slope of the CPS phase immediately gives the EPL.

\begin{figure}
    \centering
    \includegraphics[width=1.0\textwidth]{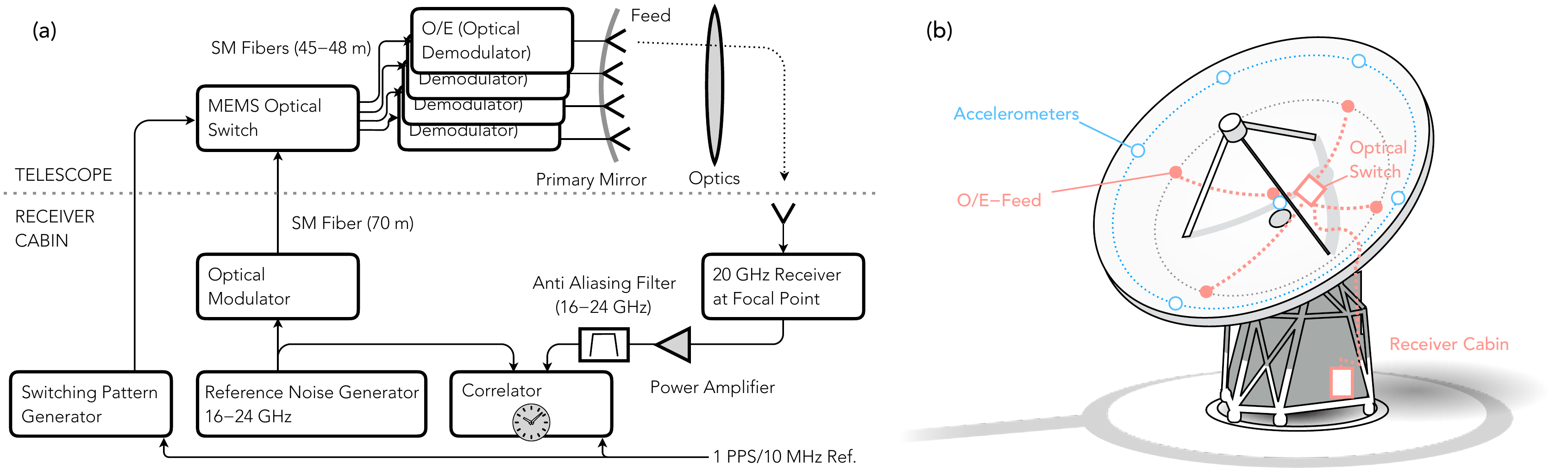}
    \caption{(a) The system block diagram of the prototype wavefront sensor for millimetric adaptive optics. (b) The configuration of the wavefront sensor system mounted on the Nobeyama 45~m telescope.}
    \label{fig:fig_system}
\end{figure}

\section{SYSTEM}
Here we describe the system of the prototype wavefront sensor for demonstration. Although this prototype only has several elements of radiators operating at 20~GHz, which is relatively low frequency and easy to handle, the aperture-plane interferometry is scalable up to tens to hundreds of elements at higher frequencies.

\subsection{Requirement and specifications}
The top-level requirement on the system is to instantaneously measure the deviation from the ideal mirror surface with an accuracy of 40~$\mu$m r.m.s.\ with a time-resolution of 100~ms, well below the natural frequency of the primary mirror structure ($\approx 1$~Hz). This requirement corresponds to the phase accuracy of 1~deg r.m.s.\ for an operating frequency of 20~GHz. The error budget is split into (1) the statistical error of thermal noises arising from the reference source and the receiver noise (appendix~\ref{sect:appendix}) and (2) the systematic error due to short term ($\sim 10$~s) stability of EPL measurements, whereas long term stability needs to be taken into account in the actual implementation of the system.
Table~\ref{tab:spec} summarizes the specifications of the prototype wavefront sensor.

\begin{table}[h]
  \begin{center}
  \caption{Specifications of the prototype wavefront sensor.}\label{tab:spec}
  \begin{tabular}{cc}
    \hline
    \hline
    Item & Value \\
    \hline
    Frequency (GHz)    & 16--24\\
    Number of elements & 5\\
    \hline
  \end{tabular}
  \end{center}
\end{table}

\subsection{Subsystems}
Figure~\ref{fig:fig_system} shows the schematic block diagram and configuration describing how the wavefront sensor works. The system is threefold; (1) the transmitter subsystem, which generates and transfers a reference signal, and injects it into the telescope optics from multiple positions across the primary surface; (2) the receiver subsystem, which collects the signal coming through the telescope optics and sends it to the correlator; and (3) the correlator subsystem, which differentiates the arrival time of the reference signals coming directly from the reference generator and through the telescope optics.

\begin{figure}
    \centering
    \includegraphics[width=1.0\textwidth]{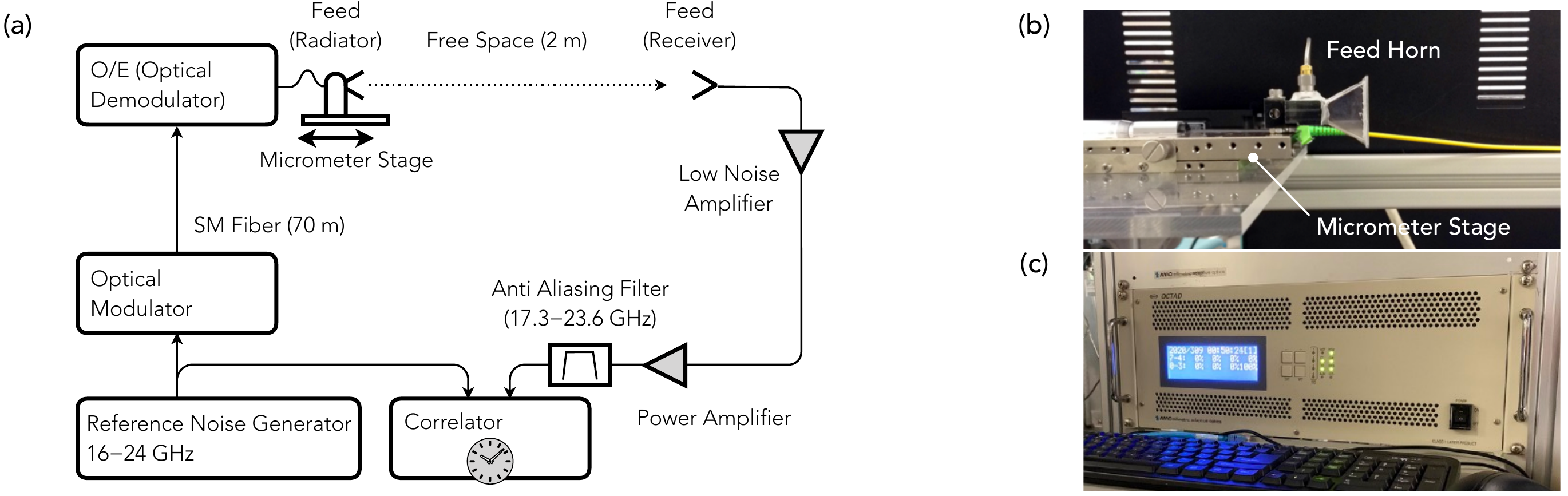}
    \caption{(a) The setup of laboratory evaluation. (b) A feed horn mounted on a micrometer stage. (c) The OCTAD-M correlator.}
    \label{fig:lab_setup}
\end{figure}

    
\subsubsection{Transmitter subsystem}
The transmitter subsystem comprises a reference noise generator, optical transmitter and switch, and radiators on the primary mirror.

\noindent\textbf{Reference noise generator.}
For the reference broadband noise, we use Johnson-Nyquist noise of a microwave terminator amplified by a cascade of power amplifiers which is followed by a bandpass filter (17.3--23.6~GHz). The excess noise ratio (ENR) is $\approx 70$~dB. The reference noise is divided, one of which goes directly into the correlator while the other is sent to the telescope primary mirror through the optical fiber.

\noindent\textbf{Optical fiber transmission and distribution.}
The reference signal is transmitted on optical fiber, since the distances to the radiators are too long to maintain the phase stability and gain of the reference signal. We use a radio-frequency on fiber (RFoF) system (Optilab, RFLL-20-H), a pair of optical modulator (E/O) and demodulator (O/E). We use a phase-stabilized single-mode fiber (Sumitomo Electric) with a linear expansion coefficient of $4.7\times 10^{-6}$~K$^{-1}$. The optical signal leaving the E/O is coupled with one of the five O/Es placed on the backup structure of the primary mirror by switching over the fibers with a MEMS-based optical switch placed closed to the central hub of the primary mirror.

\noindent\textbf{Feed horn on the primary mirror.}
The reference signal is demodulated with the O/E and is radiated by one of five identical feed horns placed on the surface of the primary mirror. We choose a linear-polarized feed horn as a reference signal radiator as the operating frequency is broad ($\Delta f/f = 0.4$). The gain and half-power beam-width at 20~GHz are 20~dBi and $10^{\circ}$, respectively. The feed horns are placed at a radius of 16~m with position angles of $-18^{\circ}$ (top), $90^{\circ}$ (left), $180^{\circ}$ (bottom), and $270^{\circ}$ (right), and at $r = 5$~m with a position angle of $0^{\circ}$ (center; see also figure~\ref{fig:fig_system}b).

\subsubsection{Receiver subsystem}
We use the H22 receiver,\footnote{https://www.nro.nao.ac.jp/\~{}nro45mrt/html/prop/status/Status\_latest.html} a science-grade cryogenic coherent receiver operating at 20~GHz. The typical system noise temperature is $T_\mathrm{sys} = 100$~K. H22 detects two circular polarization, while we only use one of them for phase measurements. The receiver output is amplified and the bandwidth is constrained by the anti-aliasing filter before the signal goes into the correlator.

\subsubsection{Correlator subsystem}

We develop a FX-type digital correlator, OCTAD-M (ELECS Inc.), which is based on the architecture initially developed for the FPGA-based fast Fourier transform (FFT) spectrometer OCTAD-S\cite{Iwai17}. OCTAD-M is equipped with two 3-bit analog-to-digital converters with 16.384~GSa/s.  We opt the third-order mode (16.384--24.576~GHz) of analog input signals, allowing us to directly sample the H22 receiver output with no down conversion. Two input signals are FFTed first and are multiplied to obtain the cross-power spectrum. The spectra are accumulated for 5 or 10~msec. OCTAD-M has a digital delay capability with up to $2^{15}$ times sampling clock time, allowing to digitally insert an instrumental delay of 0--2 $\mu$sec, corresponding to 0--599.585 m for free-space geometrical delay.

\subsubsection{Computing/control subsystem}
To synchronize the optical switch and the correlator sampling, we use common 1~PPS and 10~MHz reference signals fed by the observatory GPS server. System control and data acquisition are performed by a single Linux server.

\subsection{Ancillary devices}
To help the wavefront sensing experiment, we use ancillary devices placed on and around the Nobeyama 45~m telescope. We have six piezoelectric accelerometers attached to the backup structure of the primary mirror\cite{Hashimoto20}, which are placed close to the edge of the primary mirror with position angles of $0^{\circ}$ (top), $45^{\circ}$, $90^{\circ}$ (left), $180^{\circ}$ (bottom), $270^{\circ}$ (right), and close to the central hub (see figure~\ref{fig:fig_system}b). We also use a weather monitor placed at the top of a 50-m tall meteorological tower, which is located 75~m north of the telescope.\footnote{https://www.nro.nao.ac.jp/\~{}nro45mrt/html/obs/weather/}

\begin{figure}
    \centering
    \includegraphics[width=1.0\textwidth]{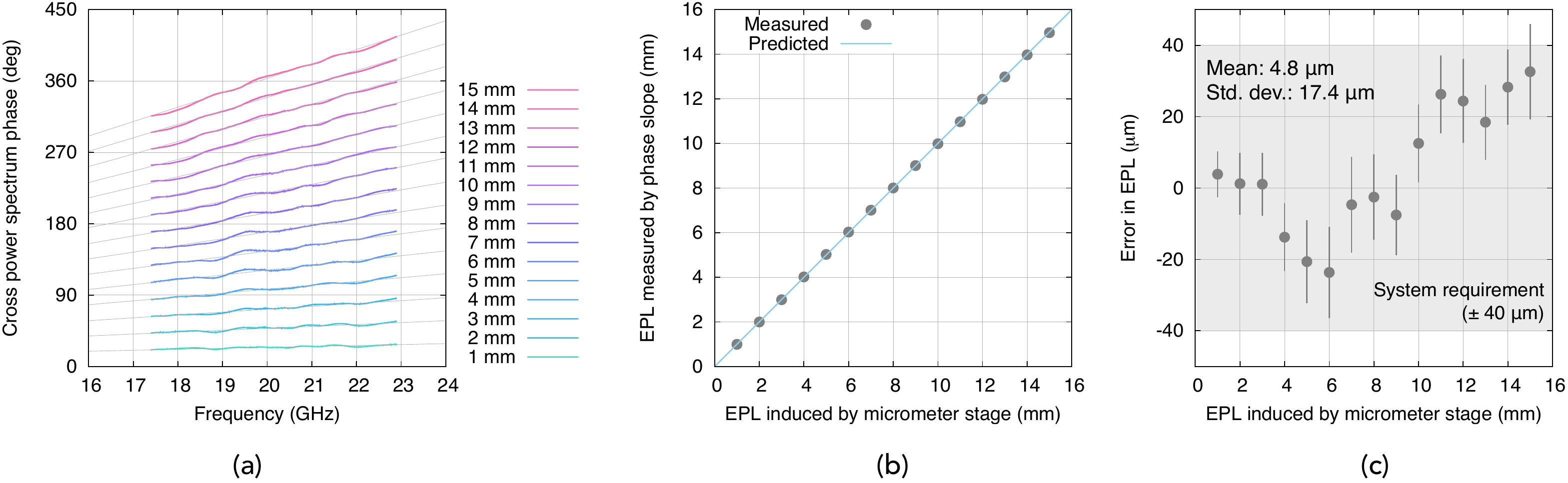}
    \caption{(a) The cross power spectrum (CPS) phases measured for different excess path lengths (EPLs) induced by mechanical deviation of the micrometer stage $z$. 
    (b) The EPL measured by the phase slope as a function of $z$. 
    (c) The residual, EPL$-z$, as a function of $z$. The error bar on each plot accounts for the $1\sigma$ error in linear regression of the CPS phase. The average of the residuals is 4.8~$\mu$m with a standard deviation of 17.4 $\mu$m. The grey region represents the system requirement ($\pm 40$~$\mu$m).}
    \label{fig:phase_stability}
\end{figure}

\section{DEMONSTRATION}

\subsection{Laboratory evaluation of linearity and accuracy in excess path length measurements}
 In advance to the on-site demonstration, we evaluate the linearity, accuracy, and stability of the system with respect to mechanical change in EPL. Although neither the actual telescope optics nor the H22 receiver is available in the lab, we use almost the same configuration of the other subsystems which are used for the on-site measurements.
The configuration of the lab experiment is shown in figure~\ref{fig:lab_setup}. Here we use a feed horn followed by a power amplifier and a bandpass filter instead of the H22 receiver. We placed the two feed horns of the radiator and receiver $\approx 2$~m away from each other. The radiator feed horn is placed on the micrometer stage that slides on a straight rail to mechanically change the EPL between the radiator and receiver, mimicking the deformation of the telescope optics. The mechanical accuracy of the micrometer stage is 3~$\mu$m.

We measure the CPS phase at $z = 0$, 1, 2, ..., 15~mm, where $z$ is the EPL induced by the mechanical deviation of the micrometer stage. The CPS is calibrated for the complex bandpass (i.e., amplitude and phase) by dividing the CPS by that of $z = 0$.  Figure~\ref{fig:phase_stability}a shows the calibrated CPS phases obtained at $z = 1$, 2, 3, ..., 15~mm. The CPS is running-averaged with a rectangle window of 300~MHz. We see a linear slope for each CPS phase although a small fluctuation remains around the best-fitting line. Since the statistical error of the CPS phase is almost negligible, the fluctuation could be dominated by systematic errors such as standing waves, multi-path interference, or phase drift of the system. 
As shown in figure~\ref{fig:phase_stability}b, the EPL derived from equation~\ref{eq:epl} is linear with respect to the mechanical deviation $z$. Figure~\ref{fig:phase_stability}c shows the residual of the measured EPL subtracted by $z$ as a function of $z$. The average of the residuals is 4.8~$\mu$m with a standard deviation of 17.4~$\mu$m, which is well below the system requirement of $< 40$~$\mu$m, confirming the system linearity and accuracy.


\begin{figure}
    \centering
    \includegraphics[width=1.0\textwidth]{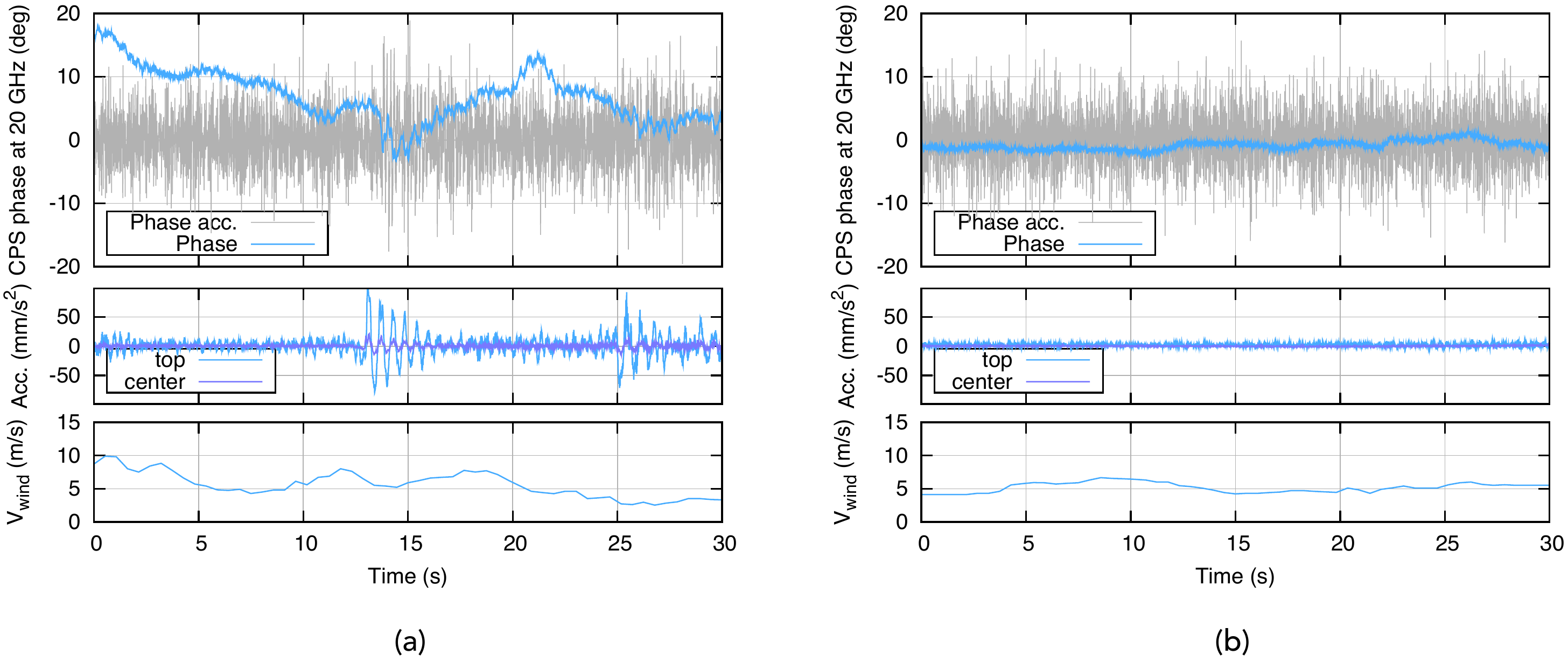}
    \caption{(a) The cross power spectrum (CPS) phase obtained under a very windy (5--10 m~s$^{-1}$) condition and (b) under a moderately windy ($\sim 5$~m~s$^{-1}$) condition. The time origins are 3:21:00 on 2020 November 22 UTC and 2:09:00 on 2020 November 23 UTC, respectively.
    (Top) The CPS phase measured at the `top' position of the 45~m primary mirror (blue curve). As the CPS is measured at 20~GHz ($\lambda \approx 15$~mm), a phase shift of 10~deg approximately corresponds to a surface deviation of 400~$\mu$m. The grey curve represents its second-order differential in arbitrary units. 
    (Center) The acceleration measured with two accelerometers attached to the top edge (blue) and the center (purple) of the 45~m primary mirror.
    (Bottom) The wind speed measured with a weather monitor placed atop the 50-m tall tower located 75~m north of the 45~m telescope.}
    \label{fig:phase_stability}
\end{figure}

 
\subsection{Demonstration at the Nobeyama 45 m telescope}

We carried out a commissioning campaign of the MAO wavefront sensor with the Nobeyama 45~m telescope in 2020 November. The goals are to confirm that the wavefront sensor actually works as designed and to demonstrate that aperture plane interferometry allows us to measure an EPL in real time. In this campaign, only 2 radiators were installed at the `top' and `center' positions. 

The top pannels of figures~\ref{fig:phase_stability}a and \ref{fig:phase_stability}b show the temporal changes in CPS phase at 20~GHz, which were taken under very windy (5--10 m~s$^{-1}$) and moderate ($\sim 5$~m~s$^{-1}$) conditions, respectively.  A large phase drift and small $\sim 3$~Hz ripples in it are clearly seen in the very windy condition, while the drift is rather suppressed under the moderate condition. This fact suggests that the CPS phase measures the EPL change induced by the wind load, although further analyses are necessary.  This is also supported by the measurements of the piezoelectric accelerometer at the `top' position; a constant $\sim 3$~Hz oscillation in addition to intermittent slow ($\sim 1.5$~Hz), high-amplitude oscillations due to a series of gust are evident in figure~\ref{fig:phase_stability}a, whereas no clear oscillation is seen in figure~\ref{fig:phase_stability}b.

\section{CONCLUSIONS}
We present a concept of a millimeter wavefront sensor which allows real-time sensing of the primary mirror surface of a future 50-m class mm/submm telescope, such as LST and AtLAST. To establish millimetric adaptive optics (MAO) that instantaneously corrects the wavefront degradation induced by deformation of telescope optics due to wind and thermal loads, our wavefront sensor employs aperture-plane interferometry to measure real-time changes in EPLs from characteristic points on the primary surface to the focal plain.
The proposed wavefront sensor operates at 16--24~GHz, which is cost-effective and is accurate enough to measure the EPL down to the level of 40~$\mu$m r.m.s.\ as demonstrated in the laboratory evaluation. Although the verification is in progress, we have demonstrated that a 2-element pilot wavefront sensor with the Nobeyama 45~m telescope worked properly and detected the EPL change induced by wind load on the telescope structure.

\appendix    
\section{SENSITIVITY}\label{sect:appendix}
In general, the variance of the cross power spectrum is given by
\[
\sigma ^2 = \frac {(S_1 + N_1) \cdot (S_2 + N_2) + S_1S_2} {2 \Delta \nu \, \Delta t},
\]
where $S_i$ and $N_i$ ($i = 1, \, 2$) are the correlated and non-correlated components of the cross power spectrum of two incident signals, respectively. Therefore, the S/N of the cross power spectrum is
\[
\mathrm{SNR} = \frac {\sqrt {S_1 S_2} \sqrt {2 \Delta \nu \, \Delta t}} {\sqrt {(S_1 + N_1) \cdot (S_2 + N_2) + S_1 S_2}}.
\]
Here, the correlated and non-correlated components input directly to the correlator from the reference signal source (reference continuum source) are $S_1, \, N_1$, and components input to the correlator via the transmission system and optical system are $S_2, \, N_2$, then $S_1 \gg N_1$, so the S/N per spectral channel of the cross power spectrum can be approximated as
\[
\mathrm{SNR} \approx \frac{\sqrt {2 \Delta \nu \, \Delta t}} {\sqrt {N_2 / S_2 + 2}} \sim \sqrt {2 \Delta \nu \, \Delta t \, (S_2 / N_2)}
\approx 8.16 \left (\frac {\Delta \nu} {\rm 1~MHz} \right)^{0.5} \left (\frac {\Delta t} {\rm 10~ms} \right) ^ {0.5} \left (\frac {N_2 / S_2} {300} \right)^{-0.5},
\]
where $N_2 / S_2 \sim O (10 ^ 2) \gg 2$ is assumed for the second approximation. If we assume that S/N of signal estimated using all $n_\mathrm{s}$ spectral channels is scaled with $\sqrt{n_\mathrm{s}}$ for S/N per channel, then
\[
\mathrm{SNR_{tot}}
\approx 739 \left (\frac{\Delta \nu}{\rm 8192~MHz} \right)^{0.5}
\left(\frac{\Delta t} {\rm 10~ms} \right)^{0.5} \left(\frac{N_2/S_2}{300}\right)^{-0.5}.
\]

Assuming that thermal noise dominates $N_2$ and the cross power spectrum has isotropic dispersion on the complex plane, the uncertainty of the phase $\theta$ is $\sigma_{\theta} \sim 1 / \mathrm{SNR_{tot}}$ (radian). Therefore,
\[
\sigma_{\theta} \approx 0.078 \ \mathrm{(deg)} \times \left (\frac{\Delta \nu} {\rm 8192~MHz} \right) ^{-0.5} \left (\frac {\Delta t} {\rm 10~ms} \right)^{-0.5} \left (\frac{N_2 / S_2}{300} \right)^{0.5}
\]
If the value is fiducial, sufficient phase measurement accuracy can be achieved. In other words, it is necessary to obtain a correlator input signal that satisfies $N_2 / S_2 <300$.

\acknowledgments 
 
We acknowledge K.\ Handa, C.\ Miyazawa, T.\ Kanzawa, T.\ Wada, and M.\ Saito for their support.
This work was supported by JSPS KAKENHI (Grant No.\ 17H06206) and NAOJ Research Coordination Committee, NINS.

\bibliography{main} 
\bibliographystyle{spiebib} 

\end{document}